# ON THE ORIGIN OF SWITCHBACKS OBSERVED IN THE SOLAR WIND

By F.S. Mozer[1,2], S.D. Bale[1,2], J.W. Bonnell[1], J.F. Drake[3], E.L.M. Hanson[1], and M.C. Mozer[4]

[1]Space Sciences Laboratory, University of California, Berkeley, CA. USA
[2]Physics Department, University of California, Berkeley, CA., USA
[3]University of Maryland, College Park, Maryland, USA
[4]Google Research, Brain Team, Mountain View, CA. USA

I ABSTRACT

The origin of switchbacks in the solar wind is discussed in two classes of theory that differ in the location of the source being either near the transition region near the Sun or in the solar wind, itself. The two classes of theory differ in their predictions of the switchback rate (the number of switchbacks observed per hour) as a function of distance from the Sun. To distinguish between these theories, one-hour averages of Parker Solar Probe data were averaged over five orbits to find:

1. The hourly averaged switchback rate was independent of distance from the Sun.
2. The average switchback rate increased with solar wind speed.
3. The switchback size perpendicular to the flow increased as R, the distance from the Sun, while the radial size increased as $R^2$, resulting in an increasing switchback aspect ratio with distance from the Sun.
4. The hourly averaged and maximum switchback rotation angles did not depend on the solar wind speed or distance from the Sun.

These results are consistent with switchback formation in the transition region because their increase of tangential size with radius compensates for the radial falloff of their equatorial density to produce switchback rates that are independent of radial distance. This constant switchback rate is inconsistent with an in situ source. The switchback size and aspect ratio, but not their hourly average or maximum rotation angle, increased with radial distance to 100 solar radii. Additionally, quiet intervals between switchback patches occurred at the lowest solar wind speeds.

II INTRODUCTION

Switchbacks are rotations of the magnetic field in the solar wind (Yamauchi et al, 2004; Landi et al, 2005, 2006; Suess, 2007; Neugebauer et al, 2013; Matteini et al, 2014; Borovsky, 2016). They have been observed in abundance on the Parker Solar Probe and they have been described extensively (Bale et al, 2019; Kasper et al, 2019; Dudok de Wit, 2020; Krasnoselskh et al, 2020; Horbury et al, 2020; McManus et al, 2020; Mozer et al, 2020; Tenerani, A. et al, 2020).

Two classes of theory attempt to explain the origin of switchbacks. In one class of theory, processes in the transition region, such as magnetic field reconnection, field line stirring, nanoflares, etc., eject switchbacks that propagate into the solar wind (Axford and McKenzie, 1992, 1997; Fisk, 2005; Du Pontieu et al, 2009; Samanta et al, 2019; Fisk and Kasper, 2020; Drake et al, 2020; Magyar et al, 2021a, 2021b). In the other class of theory, processes in the solar wind itself, such as strong turbulence, excitation of the Kelvin-Helmholtz instability, or magnetic field expansion, produce the switchbacks (Malagoli et al, 1996; Baty et al, 2002; DeForest, et al, 2016, 2018; Chhiber et al, 2018; Ruffolo et al, 2020; Squire et al, 2020). Because the radial dependence of the switchback rate differs in the two classes of theory, this dependence was investigated in order to distinguish between the transition region and in-situ models for the switchback origin.

III DEFINITION OF A SWITCHBACK

Figure 1 presents three components of the magnetic field during 25 days near perihelion three, as measured in the spiral coordinate system. In this coordinate system, which is used for all data in this paper, the X-direction is perpendicular to the Parker spiral in the ecliptic plane and points in the direction of solar rotation (against the ram direction), Y is perpendicular to the ecliptic plane and points southward, and Z points inward along the Parker spiral. The Parker spiral for this data was computed for a solar wind speed of 400 km/sec. The spiky structures in figure 1 indicate the possible locations of the switchbacks, as defined below, whose radial occurrence frequency will be determined.

The term 'switchbacks' has been defined in a variety of ways such that the number of switchbacks in a given time interval can vary by more than an order-of-magnitude, depending on the definition. This variation is unimportant for the present purposes because the switchback rate, and not the number of switchbacks, is the topic of discussion. Switchbacks are defined in this paper as rotations of the magnetic field from the Parker spiral through more than 90 degrees. They must also have durations greater than six seconds in order to eliminate signals from dust, time domain structures, telemetry noise, low frequency waves, etc. To complete the definition, it is necessary to define the cutoff angle such that, when the magnetic field rotation decreases below this angle, the switchback has ended. In figure 2, the number of switchbacks in the data of figure 1 are plotted as a function of this cutoff angle. For a cutoff angle near 90 degrees, unwanted, small excursions of substructures of field rotation above and below 90 degrees are captured. For a cutoff angle near 15 degrees, there are too few switchbacks for statistical significance. A switchback cutoff angle of 60 degrees was used to analyze the switchbacks in this paper. (A cutoff angle of 75 degrees was tested on subsets of the data with the result that the switchback numbers were greater for a 75-degree cutoff than for a 60-degree cutoff (see figure 2), but the slopes versus other quantities were the same as those described below for the 60-degree cutoff.)

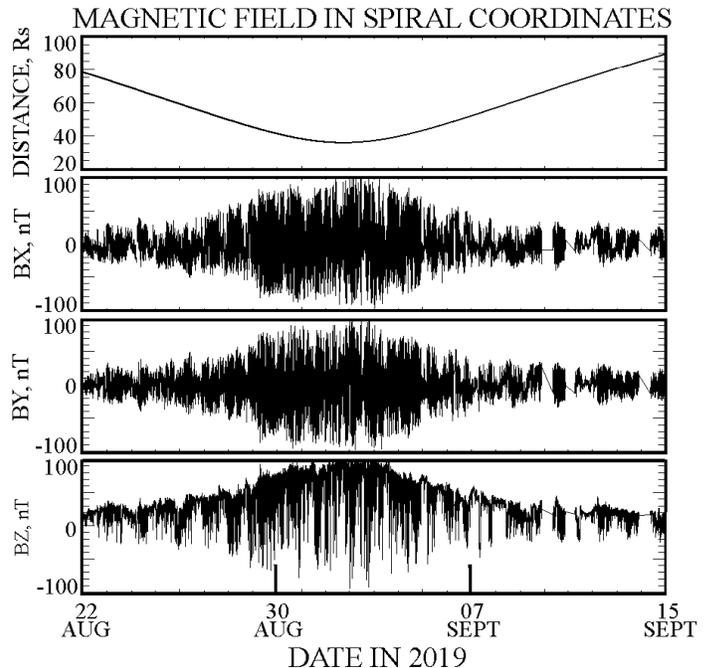

Figure 1. Magnetic field components measured in the spiral coordinate system during the perihelion of orbit three of the Parker Solar Probe.

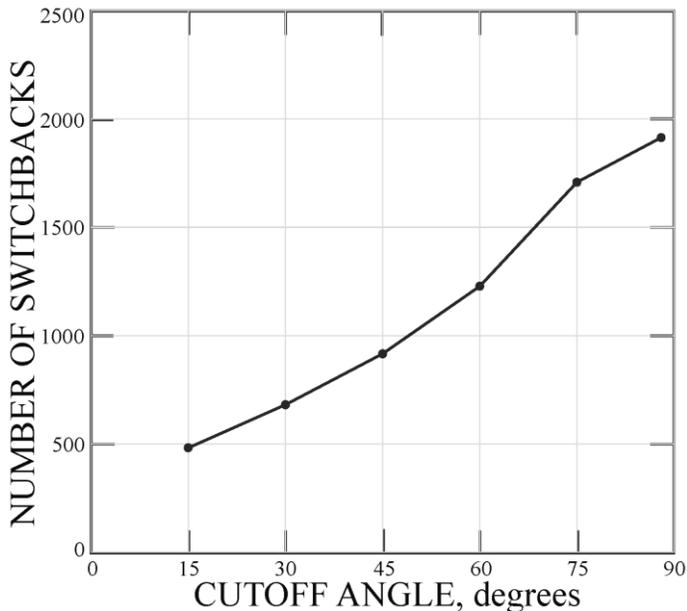

Figure 2. Number of switchbacks during the 28 days of figure 1 as a function of the cutoff angle, which is the rotation angle of the magnetic field from the Parker spiral that defines the end of a switchback.

IV DATA ANALYSIS

A two-minute interval of magnetic field rotation angles measured during an extremely active period is presented in figure 3. The average number of switchbacks per hour will be shown to be about two and this interval has four switchbacks in two minutes. The black circles at the top of the figure identify these switchback start times. For a cutoff angle near 90 degrees, the number of switchbacks can be estimated by eye to be about 12, while, for a cutoff angle of 15 degrees, there is only one switchback during this time interval. Switchbacks may also be defined as abrupt changes of the magnetic field direction. According to this definition, examination by eye suggests that there may be four or five switchbacks in the interval shown

Because the switchback rate may be a function of both the solar wind speed and the spacecraft radial position, an approach was adopted to separate the effects of the two variables and determine the switchback rate dependence on each. To improve the statistics of the analysis, the >3000 one-hour average steps at radial distances to 100 solar radii for orbits 3, 4, 5, 6, and 7 were aggregated into the data set after current sheet intervals were removed. The magnetic field data in these analyses were produced by the Fields instrument suite on the Parker Solar Probe (Bale et al, 2016) while the wind speed came from the Sweap instrument suite {Kasper et al, 2016).

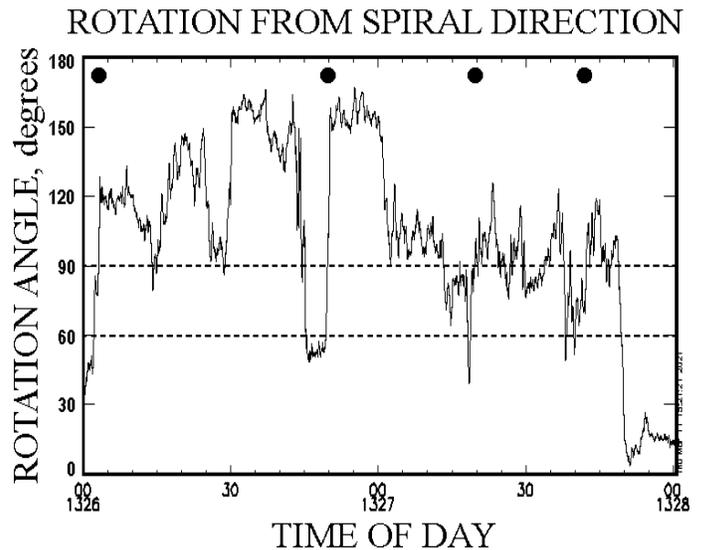

Figure 3. Two minutes of the rotation angle of the magnetic field and the resulting switchbacks.

A series of Poisson regression models was developed to predict the switchback rate as functions of wind speed and solar distance (see the Appendix for details). These models assume that the logarithm of the observed switchback rate, $s$, is a Poisson random variable with

$$\ln \mathbb{E}[s|v,d] = f(v,d)$$

where $v$ and $d$ are the solar wind speed and distance, respectively, and the expectation $\mathbb{E}[.]$ is the rate parameter of the Poisson distribution. The logarithm of the expectation was used to prevent negative values of the switchback rate from occurring. For the function $f$, polynomial functions of order $k$ with $k$ ranging from 0 to 4 were tested, where the $k'$th order function includes all terms up to order $k$, i.e.,

$$f_0(v,d) = \alpha_0$$
$$f_1(v,d) = \alpha_0 + \alpha_1 v + \alpha_2 d$$
$$f_2(v,d) = \alpha_0 + \alpha_1 v + \alpha_2 d + \alpha_3 v^2 + \alpha_4 d^2 + \alpha_5 vd$$

Via cross validation, the order $k > 1$ models did not provide better fits than the first-order model to the data on switchback rate versus solar wind speed and radial location (see the Appendix for details). The first-order model prediction is shown in Figure 4, with the color contours indicating switchbacks/hour as a function of solar wind speed and radial distance. The faint black points represent the individual data points used to obtain the fit. This plot may be understood by imagining a horizontal line at any radial distance. Along such a line, the switchback rate increases with solar wind speed. Similarly, along a vertical line at any solar wind speed, the switchback rate is approximately constant, showing that the switchback rate does not depend on radial distance.

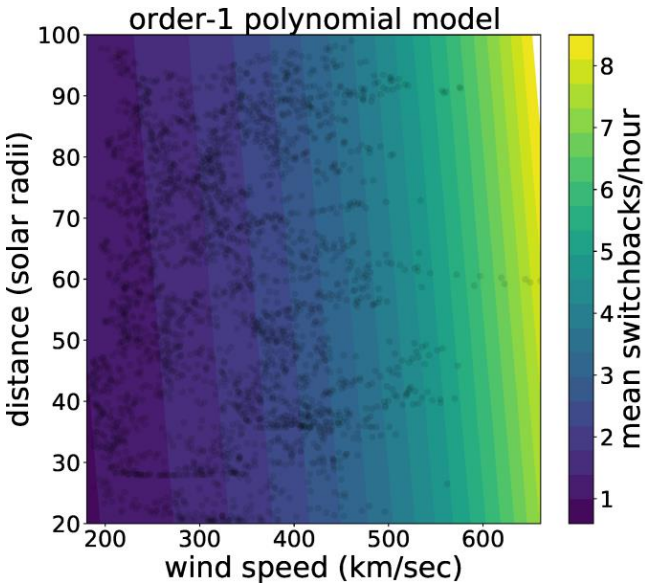

Figure 4. Poisson linear regression of the switchback rate on wind speed and solar distance. The contour lines indicating a constant switchback rate are approximately vertical, suggesting that the switchback rate increases with wind speed but does not depend on distance from the sun.

Because the first-order model linearly combines solar wind speed and radial distance, one is justified in examining the marginal regression model based only on solar wind speed, $f(v) = \alpha_0 + \alpha_1 v$, as in figure 5A. The dashed line indicates the resulting switchback rate as a function of solar wind speed and the colors of the data points indicate their radial locations. (The dashed line is slightly curved due to the nonlinearity of the link function in a Poisson regression.) The increase in switchback rate with solar wind speed is due in part, to the fact that, at higher wind speeds, more space is surveyed by the spacecraft in each hour of switchback data. At a speed of 200 km/sec the switchback rate was about one per hour in figure 5A. Thus, at 600 km/sec, (three times the speed) the switchback rate due to this effect is three switchbacks/hr. This dependence is shown by the solid curve in figure 5A, which accounts for about half of the observed switchback increase with solar wind speed.

Similarly, the marginal regression on radial distance, with $f(r) = \alpha_0 + \alpha_1 r$, is presented in figure 5B, with the dashed curve showing that the switchback rate did not depend on distance from the Sun Thus, figure 5 shows that the average switchback rate increases with solar wind speed and does not depend on radial location. It is noted that there are not many points at wind speeds greater than 600 km/sec in this data set because the spacecraft encountered mostly slow solar winds during these measurements.

To show that the results of switchback dependence on wind speed and radial position are robust, the above analysis was repeated for each of the five orbits in the data set, with the results presented in figure 6. Clearly, both the variation of switchback rate with wind speed and the lack of dependence on distance from the Sun are seen on every orbit.

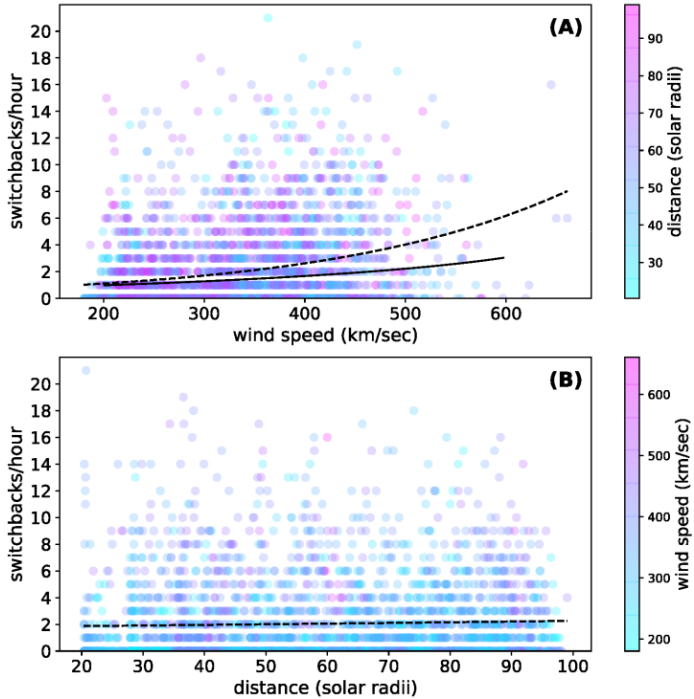

Figure 5. The dashed curves are Poisson linear regressions of the switchback rate on solar wind speed (panel 5A) and the radial location of the spacecraft (panel 5B). The panels show that the average switchback rate (the dashed curves) increases with solar wind speed and does not depend on the spacecraft radial position. The solid black curve in panel 5A is the contribution to the switchback rate resulting from the spacecraft sampling more radial information per hour at higher solar wind speeds. The colored points are the underlying observations, with the color indicating the radial position (panel 5A) or wind speed (panel 5B).

A similar analysis was performed on the maximum switchback rotation angle in each hour of data. For this analysis, cross validation indicates that first- and higher-order polynomial models predict the data no more reliably than a zeroth order (constant) model (see the Appendix for details). Figures 7 and 8 present joint and marginal first-order fits to the hourly averaged maximum rotation angle, respectively, offering little indication of a linear trend. The maximum switchback rotation angle is thus approximately independent of the solar wind speed and radial distance. The hourly averaged rotation angle is also independent of radial distance, as shown by a similar analysis.

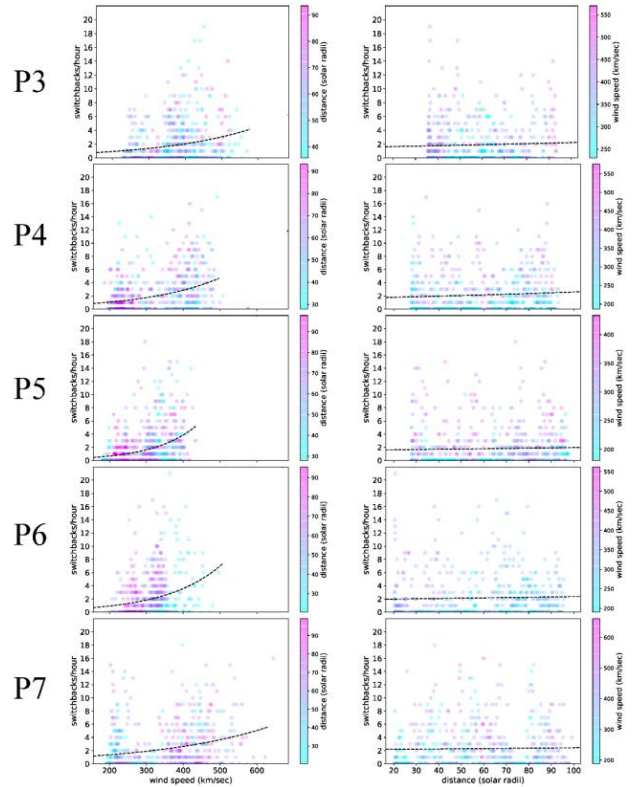

Figure 6. The switchback rate dependence on wind speed and distance from the Sun for the five Parker Solar Probe orbits studied in this paper. In every orbit, the switchback rate increased with solar wind speed and did not depend on radial position. This shows the robustness of these results.

To understand the spread of the data points in figures 5 and 8, it is important to realize that these figures are not snapshots in time. There was up to a 28-day difference in the time between points in each of the five perihelion passes and the five passes were separated by about 16 months. In addition, each pass covered a Carrington longitude of ~90 degrees. Thus, the data encompass a huge range of initial conditions on the Sun that produce a huge range of switchback numbers and sizes, solar wind speeds, etc. The curves of figures 5 and 8 provide averages over all of these conditions.

There is a suggestion in figures 4 and 5 that the number of hours without switchbacks is greatest at the lowest wind speeds. To test this possibility, figure 9 gives the percentage of hours during which the number of switchbacks was zero or one versus the solar wind speed. At the lowest speeds, there were fewer than two switchbacks in each hour about 94% of the time. This shows that the quiet intervals between switchback patches occurred at the lowest solar wind speeds.

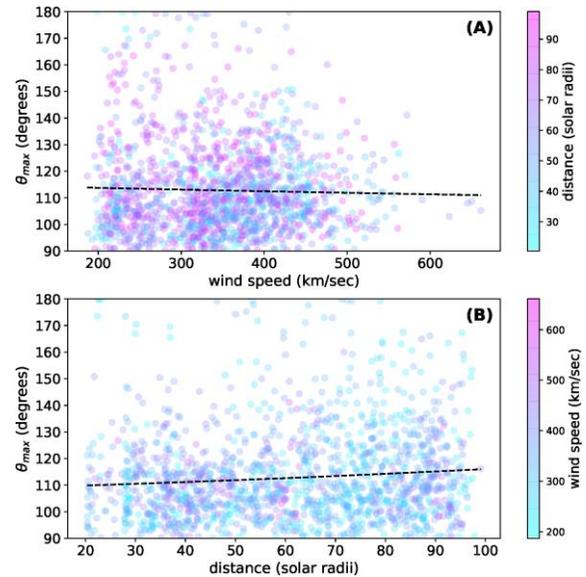

Figure 8. The dashed curves are Poisson linear regressions of the maximum angle of the switchback rotation ($\theta_{max}$) on solar wind speed (panel 8A) and the radial location of the spacecraft (panel 8B). They show that $\theta_{max}$ is approximately independent of both the spacecraft radial position and the solar wind speed. The colored points are the underlying observations, with the color indicating the radial position (panel 8A) or wind speed (panel 8B).

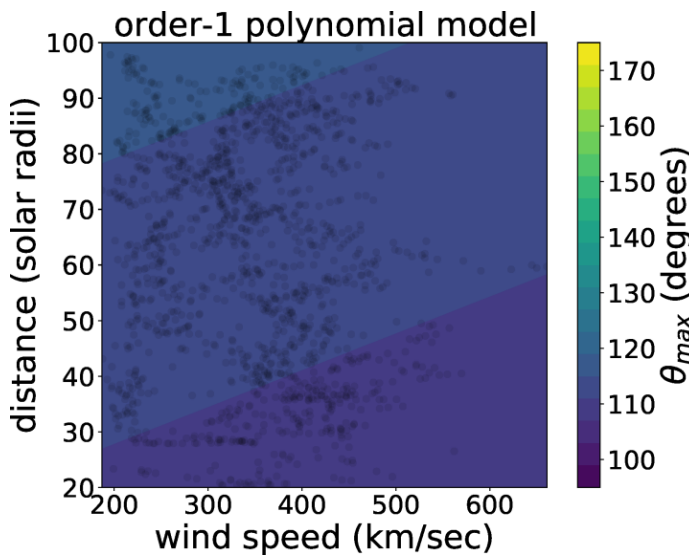

Figure 7. Poisson linear regression of the maximum rotation angle in the hourly switchback observations ($\theta_{max}$) on wind speed and solar distance. The near absence of contour lines indicates that $\theta_{max}$ does not depend on wind speed or distance from the Sun.

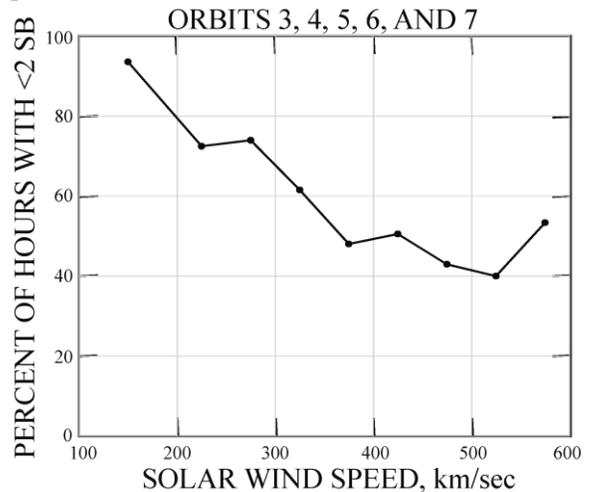

Figure 9. The percentage of one-hour intervals having zero or one switchback versus the solar wind speed. The quiet intervals with few switchbacks occur when the solar wind speed is low. The upturn of the curve at the highest speed probably results from the sparsity of data at the highest speeds.

V DISCUSSION

Before comparing switchback observation rates with the two classes of source theory (transition region and in situ sources), it is necessary to consider the size variation of switchbacks with distance from the Sun because the switchback size affects the probability of observing a switchback. Because switchbacks are magnetic structures embedded in the background magnetic field, the switchback size should increase as the magnetic field decreases with distance from the Sun. As is both expected and observed in figure 10A, the radial magnetic field decreased as $1/R^2$ while the tangential field of figure 10B decreased as $1/R$, where R is the distance from the Sun. Thus, the observed switchback duration, which is proportional to its radial dimension, should increase as $R^2$ while the tangential dimension, which is proportional to the probability of detection, should increase as R. It is not possible to measure the tangential (perpendicular to the flow) component of the switchback size on a single spacecraft, but it is possible to test the expected $R^2$ dependence of its radial dimension by measuring the switchback durations as a function of radius. In figure 10C, the quantity $D*(V/400)*(2500/R^2)$ is plotted versus R, where D is the measured duration of each switchback, V is the switchback speed and (V/400) normalizes the duration for the effect of different speeds. From the least squares solid curve, it is seen that this quantity is essentially constant with distance in figure 10C such that the radial dimension of switchbacks increased in size with distance in the same way that the radial magnetic field decreased. This justifies the assumption that the tangential dimension of switchbacks increased linearly with distance in the same way that the tangential field decreased. This radially decreasing tangential magnetic field decreases the density of a group of switchbacks in exactly the same way that their size increases, such that the probability of counting an existing switchback is independent of radius.

For an in situ source, the switchback rate must increase with radius because there is more space at larger distances for switchback formation. This expected increase of switchback rate for an in situ source is inconsistent with the observed constant switchback rate with distance. Thus, in situ sources are excluded by the observational data.

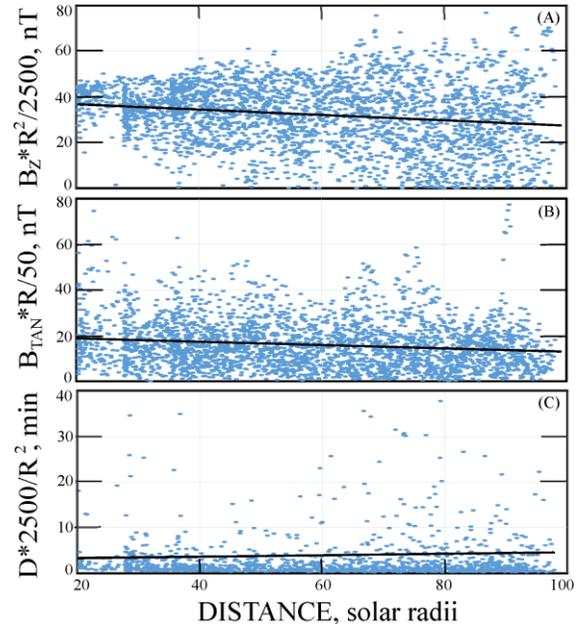

Figure 10. Panel (10A) shows that the radial component of the magnetic field decreased with distance as $1/R^2$, where R is the distance from the Sun in solar radii and the solid line is a linear least squares fit to the data. Note that the fit decreased by about 20% with distance while $R^2$ increased by a factor of 25. Panel (10B) shows that the tangential component of the magnetic field decreased as $1/R$, and panel (10C) shows, via the solid least squares fit line, that the radial extent of the switchbacks increased as $R^2$, the same rate that the radial component of the magnetic field decreased. In panel (10C), 18 data points having durations >40 minutes are not included in the graph.

On the other hand, a transition region source produces switchbacks whose counting rate is independent of radial distance outside of the transition region. Thus, the constant switchback rate with radius found in this report shows that the switchbacks were formed in the transition region near the Sun. That the number of switchbacks observed depends on the solar wind speed is as expected because switchbacks and the solar wind are generated in the same region. Because the radial magnetic field decreases with distance more rapidly than does the tangential field, the radial dimension of switchbacks increases more rapidly with distance than does the tangential dimension, resulting in switchbacks having a large and increasing aspect ratio with distance, as has been found earlier (Horbury et al, 2020; Laker et al, 2021).

The summary of this full data set is:

1. The average switchback rate did not depend on distance from the Sun.
2. The average switchback rate increased with solar wind speed.
3. The switchback tangential (to the flow) dimension increased with radial distance, R, while the radial dimension increased as $R^2$, resulting in a large and increasing switchback aspect ratio with distance from the Sun.
4. The hourly averaged and maximum switchback rotation angles did not depend on either the solar wind speed or the distance from the Sun.

These results are consistent with switchback formation at the transition region near the Sun. Once formed, their size and aspect ratio, but not their hourly averaged or maximum rotation angles, increased with radial distance to a distance of at least 100 solar radii. An additional result of this study is that quiet intervals between switchback patches occur at the lowest solar wind speeds.

VI ACKNOWLEDGEMENTS

This work was supported by NASA contracts NNN06AA01C and NASA-G-80NSSC1. The authors acknowledges the extraordinary contributions of the Parker Solar Probe spacecraft engineering team at the Applied Physics Laboratory at Johns Hopkins University. Our sincere thanks to P. Harvey, K. Goetz, and M. Pulupa for managing the spacecraft commanding and data processing, which has become a heavy load thanks to the complexity of the instruments and the orbit. The data used in this paper are available at http://fields.ssl.berkeley.edu/data

APPENDIX A

We have constructed a series of Poisson regression models to predict switchbacks-per-hour from wind speed and solar distance. These models assume that the observed integer switchback count $s$, is a Poisson random variable with

$$\ln \mathbb{E}[s|v,d] = f(v,d),$$

where $v$ and $d$ are the solar wind speed and distance, respectively, and the expectation $\mathbb{E}[.]$ is the rate parameter of the Poisson distribution. The log transform is the standard link function used for Poisson regression and ensures the rate is non-negative. For the function $f$, we tested models varying in complexity: polynomial functions of order 0 to 4, where the $k$'th order function includes all terms up to order $k$, i.e.,

$$f_0(v,d) = \alpha_0$$

$$f_1(v,d) = \alpha_0 + \alpha_1 v + \alpha_2 d$$

$$f_2(v,d) = \alpha_0 + \alpha_1 v + \alpha_2 d + \alpha_3 v^2 + \alpha_4 d^2 + \alpha_5 vd$$

etc.

The fit to the full switchback data set of models of order 1-4 is shown in Fig. A1. The coloring of the contour plot indicates the expected switchback rate. The faint black points in each graph represent the individual data points used to obtain the fit. Note the sparsity of data for high wind speeds; the fits therefore focus more on the lower wind speeds. The models of different order qualitatively agree with one another. (The fourth order model predicts very high switchback rates in the lower right quadrant of the graph, but note that there are no data in that region.)

To evaluate each model, we split the data via three rounds of five-fold cross validation. Each split yields training and evaluation data sets with 80% and 20% of the 3000 observations, respectively. The training set is used to obtain a maximum likelihood fit of the $\boldsymbol{\alpha}$ coefficients, and the evaluation set is then used to estimate two measures of model quality: the *validation score*, which is a constant offset from the log likelihood of the evaluation set under the model, and the Pearson *correlation* between the observed and predicted mean number of switchbacks.

These quality measures, shown in Figure A2 for models of order 0-4, indicate that first-order (linear) model provides the best and most parsimonious account of the data. This model justifies presenting the marginal regression curves in the main article.

We have also examined the relationship between $\theta_{max}$, the maximum rotation angle of the switchbacks in each hour of data, and the wind speed and solar distance, using an analogous procedure. Because $90° \leq \theta_{max} \leq 180°$, we used a circular regression (Sarma & Jammalamadaka, 1993) which assumes that $\theta_{max}$ is a draw from a von Mises distribution with mean $\mu = l(f(v,d))$, where we define the inverse link function $l$ as:

$$l(x) = \frac{1}{2}\tan^{-1} x + 135°.$$

Assuming an arctan function that returns a value in $[-90°, +90°]$, predictions of $\theta_{max}$ will be in the correct range with $x \in [-\infty, +\infty]$.
In agreement with the switchback prediction, our cross-validation simulations suggest that a model order 1 is appropriate for $\theta_{max}$ prediction, as shown in Figure A3. As Figure A4 indicates, models of all order have essentially the same key characteristic: little sensitivity of $\theta_{max}$ to either wind speed or distance and no strong interactions between the two variables. Figure A3 also shows this because the validation score for the k=0 model is no worse than any k≠0 model that involves wind speed and distance.

REFERENCE

Sarma, Y. R., & Jammalamadaka, S. R. (1993). Circular regression. In K. Matsusita, M. L. Puri, & T. Hayakawa (Eds.), *Statistical Science and Data Analysis: Proceedings of the Third Pacific Area Statistical Conference* (pp. 109-128). Utrecht, the Netherlands: VSP.

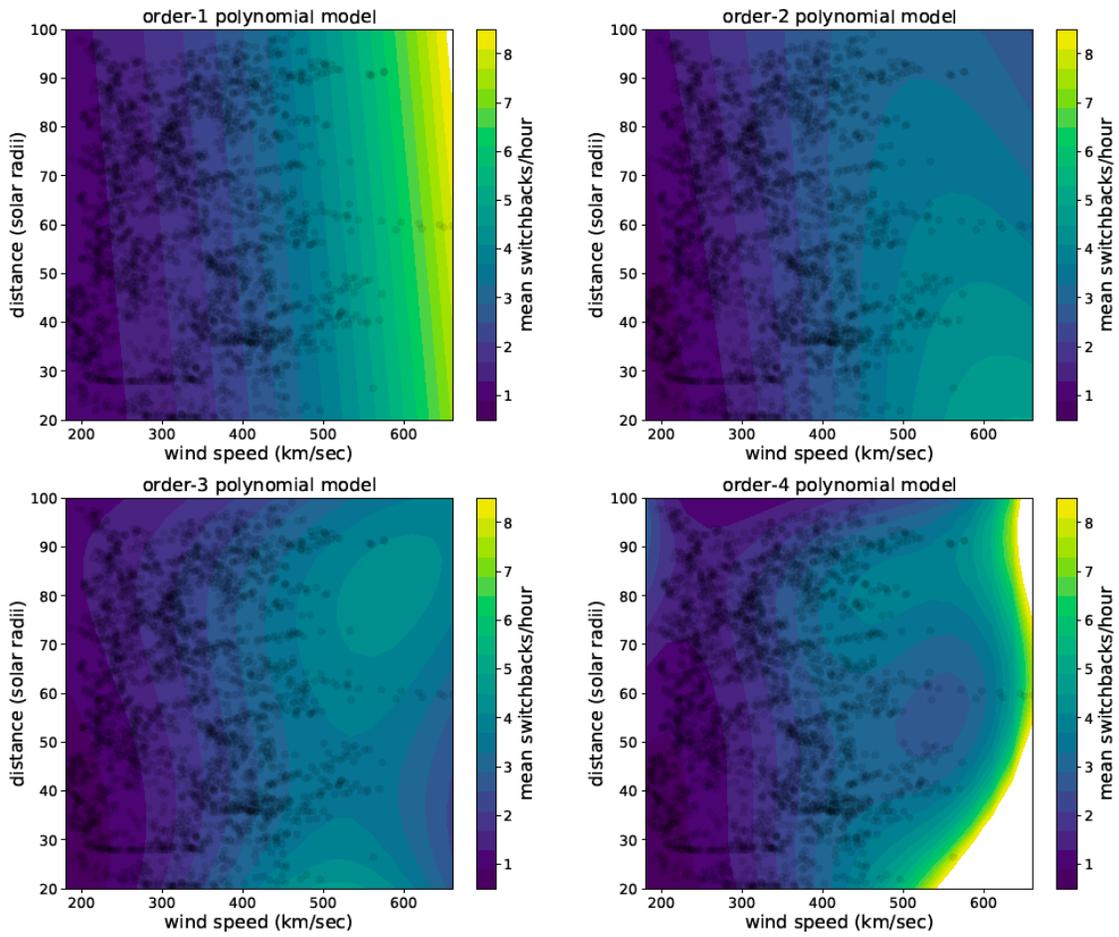

Fig. A1. The expectation (rate parameter) of switchbacks as a function of wind speed and distance, for models of order 1-4.

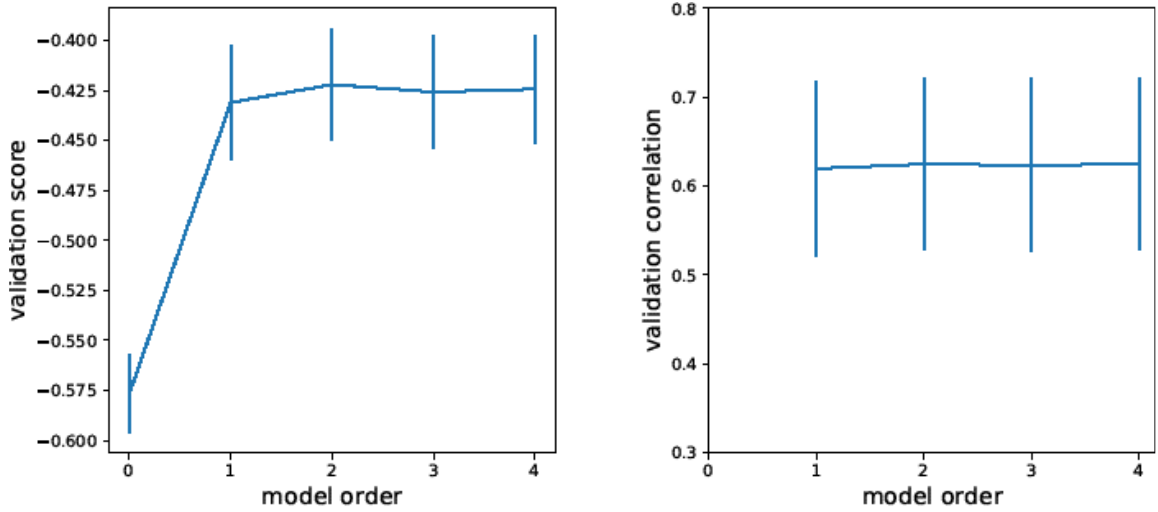

Figure A2: Validation score (related to log likelihood) and correlation are shown for models of order 0-4. The correlation is undefined for model order 0. Error bars indicate ±1 standard error of the mean over the 15 data splits.

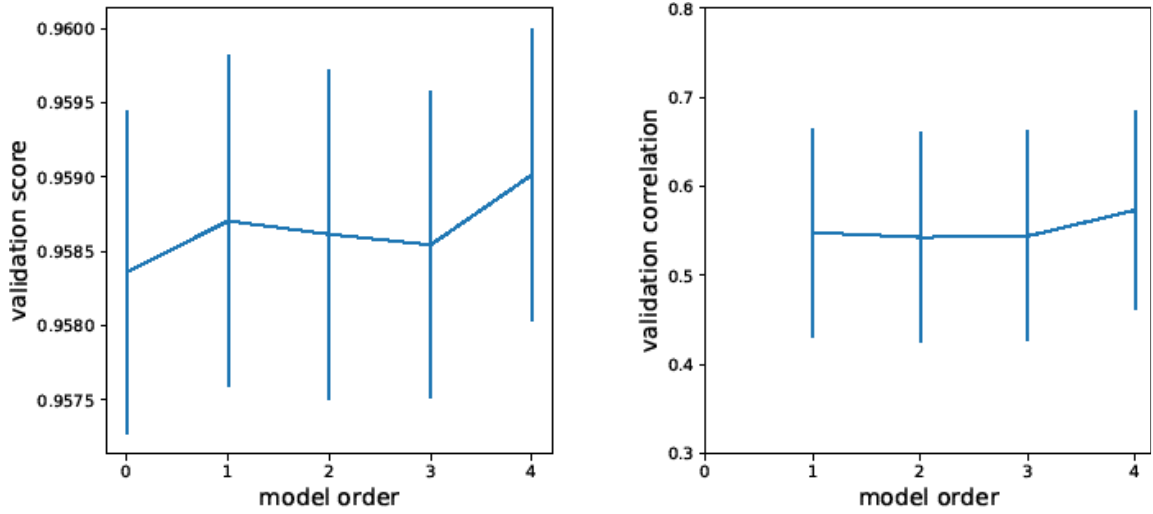

Figure A3: Validation score (related to log likelihood) and correlation are shown for models of order 0-4. The correlation is undefined for model order 0. Error bars indicate ±1 standard error of the mean over the 15 data splits.

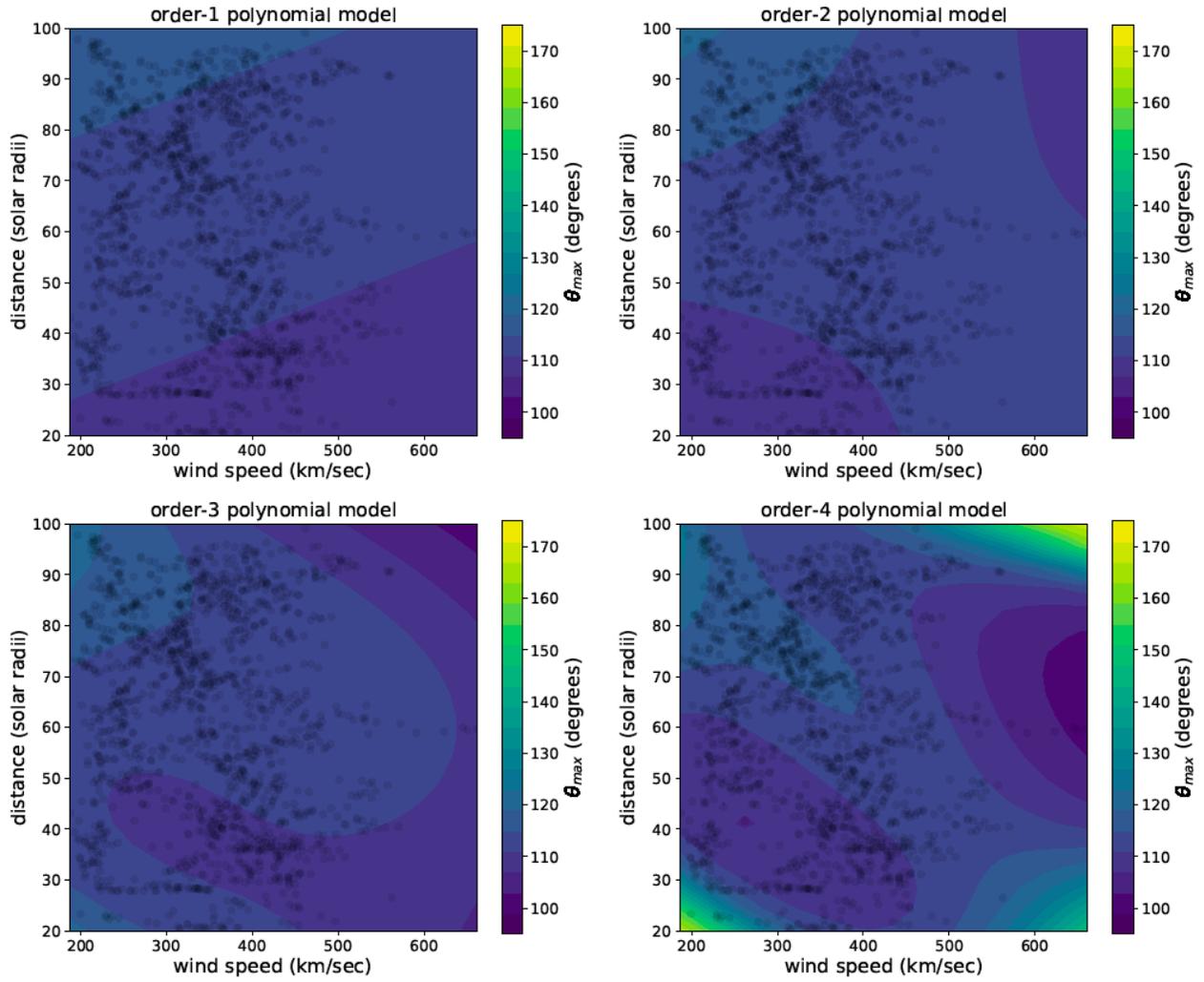

Figure A4. The expectation of $\theta_{max}$ as a function of wind speed and distance, for models of order 1-4.